%% file: ms.tex
\documentclass[apjl]{emulateapj}
\usepackage{natbib}

\newcommand{\ks}{km/s}
\newcommand{\mic}{$\mu$m~}

\slugcomment{To appear in ApJ Letters}

\shorttitle{Starburst galaxies in filaments}
\shortauthors{Fadda et al.}

\begin{document}


\title{Starburst galaxies in cluster-feeding filaments unveiled by Spitzer}


\author{Dario Fadda\altaffilmark{1}}
\affil{NASA Herschel Science Center, Caltech 100-22, CA 91125 Pasadena}
\email{fadda@ipac.caltech.edu}

\author{Andrea Biviano\altaffilmark{2}}
\affil{INAF -- Osservatorio Astronomico di Trieste, via G. B. Tiepolo 11, I-34143, Trieste, Italy}

\author{Francine R. Marleau and Lisa J. Storrie-Lombardi}
\affil{Spitzer Science Center, Caltech 220-6,  CA 91125 Pasadena}

\and

\author{Florence Durret}
\affil{Institut d'Astrophysique de Paris, CNRS, Universit\'e Pierre et Marie Curie, 98bis
Bd Arago, F-75014 Paris, France}


\altaffiltext{1}{Visiting Astronomer, Kitt Peak National Observatory, National Optical Astronomy Observatories, which is operated by the Association of Universities for Research in Astronomy, Inc. (AURA) under cooperative agreement with the National Science Foundation.}
\altaffiltext{2}{Visiting Astronomer, Italian Telescopio Nazionale Galileo (TNG) operated on the island of La Palma by the Fundaci\'on Galileo Galilei of the INAF (Istituto Nazionale di Astrofisica) at the Spanish Observatorio del Roque de los Muchachos of the Instituto de Astrofisica de Canarias}


\begin{abstract}
We report the first direct detection with {\em Spitzer} of galaxy
filaments.  Using {\em Spitzer} and ancillary optical data, we have
discovered two filamentary structures in the outskirts of the cluster
Abell~1763. Both filaments point toward Abell~1770 which lies at the
same redshift as Abell~1763 ($z=0.23$), at a projected distance of
$\sim$13~Mpc. The X-ray cluster emission is elongated along the same
direction.

Most of the far-infrared emission is powered by star
formation. According to the optical spectra, only one of the cluster
members is classified as an active galactic nucleus. Star formation is
clearly enhanced in galaxies along the filaments: the fraction of
starburst galaxies in the filaments is more than twice than that in other
cluster regions.  We speculate that these filaments are feeding the
cluster Abell~1763 by the infall of galaxies and galaxy
groups. Evidence for one of these groups is provided by the analysis
of galaxy kinematics in the central cluster region.

\end{abstract}


\section{Introduction}
\label{intro}
Thanks to large redshift surveys, it is presently well established
that baryonic matter is arranged in the Universe in a network of voids
and filamentary structures
\citep[e.g.][]{dLGH86,Doroshkevich01,PDH04}. Galaxy clusters are found
at the intersections between filaments. Hierarchical models of galaxy
formation \citep[e.g.][]{Benson05} and observations
\citep[e.g.][]{ZF93,Ceccarelli05,Cortese06} show that clusters
constantly accrete small galaxy groups through filaments.  These
accreted galaxies undergo an accelerated evolution when they meet the
denser cluster environment, experiencing a short starburst phase
followed by the quenching of their star formation activity as their
gas reservoirs are consumed and/or stripped off
\citep[e.g.][]{Bekki99,BNM00,Fujita04}. These events are traced by
differences in morphology and star formation rate of galaxies inside
and outside clusters
\citep[e.g.][]{Dressler80,Balogh98,Goto03,Kodama04}.

Much of what we have learned so far about the evolution of galaxies in
clusters comes from detailed studies of the inner cluster regions,
where the contrast with respect to the field is high, so that most
spectroscopically targeted galaxies have a high probability of being
cluster members. We are therefore limited to study the later stages of
the environmental processes that drive galaxy evolution in clusters.
We can follow the entire evolution of these galaxies, only observing the
external regions where infalling field galaxies first encounter the
cluster potential and the intracluster gas. Unfortunately, studies
of this kind are extremely rare since they require wide-field images
coupled with intensive spectroscopic follow-up observations
\citep[see, e.g.,][]{PR07}.

Moreover, in order to better characterize galaxy properties through
their spectral energy distributions (SED), multi-wavelength
observations are needed.  Until recently, most studies of the
external regions of galaxy clusters have been based on
optical/near-infrared data \citep[see,
e.g.,][]{Treu03,Gerken04,Rines05}. On the other hand, infrared studies
of clusters have been limited to the inner regions \citep[see the
review of][]{MFB05}. With the advent of {\em Spitzer}
\citep{Werner04}, the outskirts of galaxy clusters have now started to
be observed \citep{Bai06,Bai07,Geach06,Marcillac07}.

In this letter we report our discovery with {\em Spitzer} 
of two galaxy filaments in the outer region of the cluster Abell~1763
(at a redshift $\overline{z}=0.23$).  The two filaments are rich in actively
star-forming galaxies, which are probably exhausting their gas supply
while entering the virialized cluster environment. Throughout this
paper we use $H_0=70$ km~s$^{-1}$~Mpc$^{-1}$, $\Omega_m=0.3$,
$\Omega_{\Lambda}=0.7$.

\section{{\em Spitzer} and Ancillary Observations}
\label{data}
Abell~1763 was surveyed with the {\em InfraRed Array Camera}
\citep[IRAC][]{Fazio04} and the {\em Multiband Imaging Photometer for
Spitzer} \citep[MIPS][]{Rieke04} as part of a program to observe
three medium-redshift clusters up to large distances from the center
study the star formation activity in infalling galaxies.
 A field of 40'$\times$55' was observed with MIPS at 24,
 70, and  160~$\mu$m. An overlapping field of 39'$\times$39' was
observed at 3.6, 4.5, 5.8, and 8.0~$\mu$m with IRAC. The entire field
is covered by the Sloan Digital Sky Survey (SDSS) in the
$u',g',r',i',$ and $z'$ bands. To detect fainter objects in
preparation for the spectroscopic follow-up observations, we
obtained an $r'$ image of the 24~$\mu$m field with the {\em Large
Format Camera} on the Palomar 200in telescope, reaching two magnitudes deeper
than the SDSS image. We obtained 382 spectra with the {\em Hydra}
spectrograph on the WIYN telescope at Kitt Peak for most of the
sources brighter than 0.3~mJy at 24~$\mu$m, corresponding to an
average signal-to-noise ratio of 5, with optical counterparts
brighter than $r' =20.5$. To obtain a better description of the
velocity distribution of the cluster, we obtained additional 198 spectra
mainly in the central region with the {\em DOLORES} multi-object
spectrograph on the TNG telescope in La Palma, Spain. In this case, we
targeted fainter galaxies with $r' <21.5$ including faint infrared
sources and early-type galaxies. Finally, to obtain estimates of the stellar
mass of the galaxies as well as a better description of their SEDs, we
obtained $J, H,$ and $K_s$ images of most of the 24~$\mu$m field with
the {\em Wide InfraRed Camera} on the Palomar 200in telescope. These
images are typically 3 mags deeper than the 2MASS \citep{Skrutskie06}
images in this field.
The data and their analysis will be presented in forthcoming papers.

\section{Spatial and redshift distributions}
\label{distrib}
The number of redshifts from our WIYN and TNG observations, as well as
that from SDSS, are reported in Table~\ref{tbl:z}.  Cluster members
were identified using the procedure developed by \citet{dHK96}.
\begin{figure*}
\epsscale{.80}
\plotone{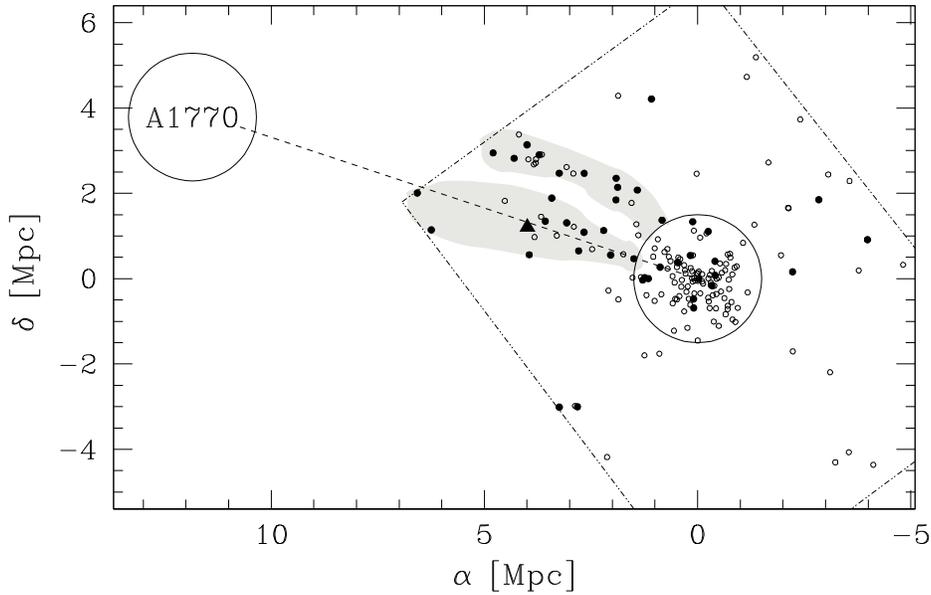}
\caption{Spatial distribution of the 181 spectroscopically selected
members of the cluster Abell~1763. Target selection was done
exclusively with 24$\mu$m sources in the region outside the 1.5~Mpc
radius central region. In the inner region, early-type galaxies
were targeted also  to better describe the velocity distribution of the
cluster.  The dotted rectangle corresponds to the 24~$\mu$m MIPS
field. Starburst galaxies and AGN are marked with filled dots and
triangles, respectively. Circles of 1.5 Mpc radius indicate the
centers of Abell~1763 and Abell~1770, connected by a dashed line.  The
shaded regions have been traced by eye to highlight two filamentary
structures. In these regions there is a clear excess of starburst
galaxies.
\label{fig1}}
\end{figure*}
The spatial distribution of the selected cluster members
(Figure~\ref{fig1}) presents a remarkable asymmetry outside $r_{500}$
($\sim 1.5$~Mpc), the radius at which the mean interior overdensity is
500 times the critical density.  We computed $r_{500}$ using the
relationship by \citet{Carlberg97} between $r_{200}$ and velocity
dispersion and a \citet{NFW97} profile with concentration $c=4$ to
obtain $r_{500}$ from $r_{200}$. The asymmetry is caused by the
presence of two galaxy filaments oriented at $\sim 51^{\circ}$ and
$\sim 73^{\circ}$ counterclockwise from the North direction, hereafter
referred to as the Northern (N) and Southern (S) filaments,
respectively. The S (and perhaps also the N) filament is directed
toward the neighboring cluster Abell~1770, located at approximately
the same redshift of Abell~1763 ($\overline{z}=0.2313$), but $\sim 13$
Mpc away. The N and S filaments have different mean velocities: $(-135
\pm 138)$ and $(300 \pm 99)$~\ks, respectively, relative to the
cluster mean velocity, suggesting that the two filaments are entering
the cluster from opposite directions (see Figure~\ref{fig2}).

Within $r_{500}$, the filaments are no longer distinguishable from the
inner cluster. However, the galaxy distribution in the inner region
and the X-ray emitting intra-cluster gas are elongated along roughly
the same direction of the two filaments. The X-ray image also shows an
emission tail south-west of the center which leads \citet{LND07} to
suggest that Abell~1763 is undergoing accretion events.  Moreover, the
X-ray temperature map conserves traces of a previous merger from the
south-east \citep{PD07}.  An ongoing accretion scenario is reinforced
by the radio observations of the brightest cluster galaxy (BCG
hereafter) which is a wide-angle tail source with bent radio emission
\citep{LND07}. The bending is probably caused by the pressure of the
intra-cluster gas as the BCG moves relative to it, heading toward
Abell~1770. Although the direction of motion is opposite to what would
be naively expected if the filaments were feeding the cluster, it can
be explained as the result of a non-dissipative collision with an
infalling group reaching the cluster from the filaments which
displaced the BCG from its original location (the galaxy is now 0.1
Mpc off the X-ray peak emission).

Additional supporting evidence for an ongoing merger event in the
cluster center comes from the analysis of the velocity distribution of
the cluster member galaxies.  In the inner region, the velocity
distribution deviates from gaussianity (see Figure~\ref{fig2}). The
value of the kurtosis, $K=2.1$, is significantly less than that
expected for a Gaussian distribution ($>99$\% confidence level). Such
a platykurtic distribution can be interpreted as the superposition of
two Gaussians with different means (see Figure~\ref{fig2}). A similar
decomposition is obtained also through a non-parametric
adaptive-kernel technique \citep{FSB98}. The two systems differ in
mean velocity by $\sim 2500$~\ks, and have velocity dispersions
$\sigma_v \sim 1100$ and $\sim 650$~\ks. These correspond to X-ray
temperatures $T_X \sim 7$ and $\sim 2.5$~keV, respectively, using the
$T_X-\sigma_v$ relationship \citep[e.g.][]{Girardi96}. These values
are close to the observational estimates by \citet{LND07},
$T_X=(7.2 \pm 0.4)$~keV for the main cluster and $T_X=(3.4 \pm
0.4)$~keV for the detected south-west emission tail. It is therefore
tempting to identify the two systems detected in X-ray with the two
systems detected in velocity space.
\hspace*{-1cm}\input{tab1.tex}

\section{Galaxy properties}
\label{props}
We have fitted the SEDs of the 181 spectroscopically identified
cluster members using a set of templates built with GRASIL
\citep{Silva98} which includes elliptical, spiral, starburst, and
post-starburst SEDs \citep{Biviano04}.  Using an altenative set of
templates \citep[][]{Polletta07} we obtained similar results.  SEDs of
active galactic nuclei (AGN) were not considered since only one
cluster member, which lies on the S filament (see Figure~\ref{fig1}),
is an AGN according to its optical spectrum. Moreover, the sources
with IRAC fluxes do not occupy the color diagram region where AGN are
expected \citep[][]{Lacy07} and none of the cluster members
corresponds to point sources in the XMM image of the field.

Using the best-fit templates we have then determined the galaxy
stellar masses ($M_{\star}$) and star-formation rates
(SFR). $M_{\star}$ are determined from the rest-frame $K_s$-band
luminosities $L_K$ by adopting $M_{\star}/L_K=1.2$ and 0.5, for early-
and late-type templates, respectively \citep{LMS03}. SFR
are determined from total infrared luminosities using the relationship of
\citet{Kennicutt98}.

We define the ratio $f_{sb} \equiv \mbox{SFR} \times \tau_{sb} /
M_{\star}$, where $\tau_{sb}$ is the timescale over which the galaxy
is assumed to form stars at the currently observed SFR. We take
$\tau_{sb}=100$~Myr as typical durations of starburst episodes
\citep{BGK00}. High values of $f_{sb}$ identify galaxies with 
elevated SFRs given their stellar mass. Hereafter, galaxies with
$f_{sb}>0.25$ are called {\em starburst galaxies}. Given that a galaxy
gas content is at most 25\% of its total baryonic mass \citep{YS91},
starburst galaxies will consume their entire gas content in the
current starburst episode, if it lasts indeed $\sim 100$ Myr.

Starburst galaxies preferentially inhabit the two filaments (see
Figure~\ref{fig1}). This is confirmed by the following statistical
analyses, which are performed on the subsample of cluster members with
24~\mic emission, in order to ensure completeness. A bidimensional
Kolmogorov-Smirnov test \citep{ff87} indicates that the spatial
distribution of starburst galaxies is indeed significantly different
from that of other cluster members, ($>99$\% confidence
level). Moreover, the fraction of starburst galaxies is $0.6 \pm 0.1$,
$0.3 \pm 0.1$, and $0.2 \pm 0.1$ in the filaments, the central
$r_{500}$ region, and the outer region excluding the filaments,
respectively. In the same regions, the average $f_{sb}$ is $0.26 \pm 0.02$,
$0.14 \pm 0.02$, and $0.17 \pm 0.02$, respectively. The distributions of
the $f_{sb}$ values are nearly identical in the two filaments.

\section{Discussion and conclusions}
\label{disc}
Using infrared {\em Spitzer} photometric data, complemented with
optical/near-infrared photometric and spectroscopic data, we have
discovered two galaxy filaments joining into the $z=0.23$ cluster
Abell~1763. The 24~\mic selection has made the filaments stand out
very clearly against the background and the dense cluster region. This
is due to the enhanced star-forming activity among the filament galaxies,
as most of the 24~\mic emission is of stellar origin \citep[see
also][]{Geach06,Marcillac07}.

\begin{figure}[t]
\plotone{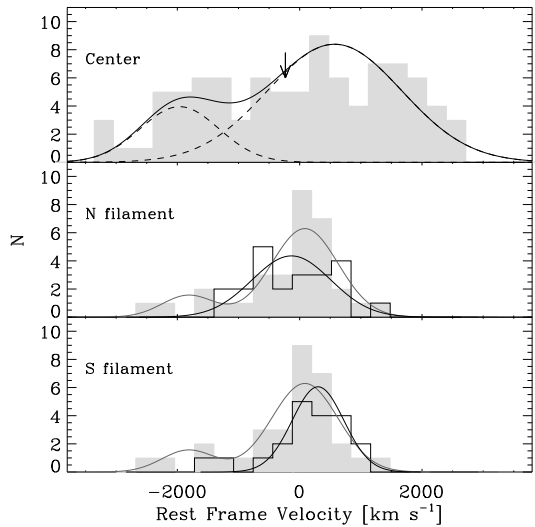}
\caption{Rest-frame velocity distributions for different cluster
regions.  {\em Upper panel}: galaxies within 1.5 Mpc of the X-ray
cluster center. The solid line represents a two-Gaussian fit to the
velocity distribution, and the dashed lines represent the two separate
components of the fit, with velocity dispersions of $641 \pm 108$
km~s$^{-1}$ (left) and $1097 \pm 152$ km~s$^{-1}$ (right).  The
vertical arrow indicates the BCG velocity. {\em Middle and lower
panels}: galaxies outside 1.5 Mpc. The shaded histogram shows the
velocity distribution of galaxies outside the filaments.  The grey
line represents a fit to this velocity distribution with two Gaussians
with velocity dispersions of $431 \pm 183$ km~s$^{-1}$ (left) and $543
\pm 104$ km~s$^{-1}$ (right).  The solid line histograms represent the
velocity distributions of galaxies in the N-filament (middle) and
S-filament (lower). The solid lines represent robust Gaussian fits to
the filament velocity distributions, with velocity dispersions of $645
\pm 92$ km~s$^{-1}$ and $422 \pm 59$ km~s$^{-1}$ (N-filament and
S-filament, respectively). Errors on the velocity dispersions are
1-$\sigma$, based on 1000 bootstrap resamplings.
\label{fig2}}
\end{figure}

Both filaments extend in the direction toward the neighboring cluster
Abell~1770, which is located $\sim 13$ Mpc away. The Abell~1763
intra-cluster gas is elongated in the same direction, as indicated by
its X-ray morphology \citep{LND07}. The cluster is probably being fed
by the two filaments along which galaxies and galaxy groups accrete
onto the main cluster. Evidence for ongoing accretion is provided by
the detection of substructures both in the X-ray emitting
intra-cluster gas \citep{LND07} and in the velocity distribution of
cluster galaxies (see Figure~\ref{fig2}).  Since the spatial
distribution of the galaxies identified as part of the low-velocity
subcluster is not compact, the subcluster-cluster collision has
probably occurred already, leading to the disruption of the colliding
subcluster. This scenario could explain the displacement of the BCG
from the cluster center, and its relative motion with respect to the
intra-cluster gas \citep[][]{LND07}.

Galaxies belonging to the two filaments have higher SFR, on average,
than other cluster members, both those in the inner $r_{500}$ region
and those outside but not in filaments. We therefore conclude that
filaments are environments favorable to the onset of a galaxy
starburst activity. Processes related to the dense intra-cluster gas
are not effective in filaments, leaving gravitational interactions
among galaxies as the most promising candidate for the stimulation of
starbursts. In fact, the relatively high density of galaxies in
filaments compared to the general field, and their relatively low
velocity dispersion ($\sim 500$~\ks) compared to the cluster enhances
the tidal effect of galaxy encounters \citep{Mamon96}, and hence the
probability of an induced star-forming activity \citep{BH96}.

Our results and conclusions find support in recent {\em Spitzer}
investigations of other clusters.  \citet{Marcillac07} have found that
the mid-infrared selected galaxies in a distant cluster ($z=0.83$) are
associated with infalling galaxies. \citet{Bai07} have suggested that
the cluster environment is able to stimulate the star-formation
activity in infalling field galaxies before they enter the cluster
central regions where gas is stripped and star-formation
suppressed. Our results are also consistent with those of
\citet{PR07}, who have used optical data to discover an enhanced
star-forming activity among galaxies associated with filaments in the
nearby Pisces-Cetus supercluster. They also claim that the SFR in the
filament galaxies peaks at 3--4 Mpc from the cluster center which is
consistent with our findings. However, we could better assess this
result by extending our coverage further to the East.

Since our results are based on the study of a single cluster, it would
be hazardous to generalize them to the entire cluster population, as
different clusters may follow different evolutionary paths. With the
completion of our observational program, another two clusters at $z
\simeq 0.2$ will be analysed to confirm that galaxy evolution is
indeed speeding up in cluster-feeding filaments.\\

\acknowledgments This work is based in part on observations made with
{\em Spitzer}, a space telescope operated by the Jet Propulsion
Laboratory, California Institute of Technology under a contract with
NASA. Support for this work was provided by NASA through an award
issued by JPL/Caltech.  Funding for the SDSS and SDSS-II has been
provided by the Alfred P. Sloan Foundation, the Participating
Institutions, the National Science Foundation, the US Department of
Energy, NASA, the Japanese Monbukagakusho, the Max Planck Society, and
the Higher Education Funding Council for England. The SDSS is managed
by the Astrophysical Research Consortium for the Participating
Institutions (see list at www.sdss.org/collaboration/credits.html).

{\it Facilities:} \facility{Spitzer (MIPS)}, \facility{Spitzer (IRAC)}, \facility{Palomar 200in (LFC)}, \facility{Palomar 200in (WIRC)}, \facility{WIYN (Hydra)}, \facility{TNG (Dolores)}.

\end{document}

%% file: tab1.tex
\begin{deluxetable}{ccccc}
\tabletypesize{\scriptsize}
\tablecaption{Redshifts measured\label{tbl:z}}
\tablewidth{0pt}
\tablehead{
\colhead{Survey} & \colhead{N$_z$} & \colhead{N$_z$} & \colhead{N$_z$} & \colhead{Flux limits} \\
&Total&Cluster&Cluster&\\ 
&&&\& 24$\mu$m&\\ 
}
\startdata
WIYN  & 360 & 80 & 80 & r'$<$20.5, f$_{24} >$0.3~mJy\\
TNG   & 157 &101 & 34 & r'$<$21.5\\
SDSS  &  96 & 18 & 11 & r'$<$20.8\\
Total & 581 &181 &106 & \\
\enddata
\end{deluxetable}

%% file: ms.bbl
\begin{thebibliography}{}
\bibitem[Bai et al.(2006)]{Bai06}
  Bai, L. et al. 2006, \apj, 639, 827

\bibitem[Bai et al.(2007)]{Bai07}
  Bai, L. et al. 2007, \apj, 664, 181

\bibitem[Balogh et al.(1998)]{Balogh98}
  Balogh, M. L. et al. 1998, \apj, 504, L75

\bibitem[Balogh et al.(2000)]{BNM00}
  Balogh, M. L., Navarro, J. F., \& Morris, S. L. 2000, \apj, 540, 113

\bibitem[Barnes \& Hernquist(1996)]{BH96}
  Barnes, J. E. \& Hernquist, L. 1996, \apj, 471, 115

\bibitem[Barton et al.(2000)]{BGK00}
  Barton, E. J., Geller, M. J., \& Kenyon, S. J. 2000, \apj, 530, 660

\bibitem[Bekki(1999)]{Bekki99}
  Bekki, K. 1999, \apj, 510, L15

\bibitem[Benson(2005)]{Benson05}
  Benson, A. J. 2005, \mnras, 358, 551

\bibitem[Biviano et al.(2004)]{Biviano04}
  Biviano, A., et al. 2004, \aap, 425, 33


\bibitem[Carlberg et al.(1997)]{Carlberg97}
  Carlberg, R. G., et al. 1997, \apj, 485, L13

\bibitem[Ceccarelli et al.(2005)]{Ceccarelli05}
  Ceccarelli, M. L., et al. 2005, \apj, 622, 853

\bibitem[Cortese et al.(2006)]{Cortese06}
  Cortese, L.,  et al. 2006, \aap, 453, 847

\bibitem[de Lapparent, Geller \& Huchra(1986)]{dLGH86}
  de Lapparent, V., Geller, M. J., \& Huchra, J. P. 1986, \apj, 302, L1

\bibitem[den Hartog \& Katgert (1996)]{dHK96}
  den Hartog, R. \& Katgert, P. 1996, \mnras, 279, 349

\bibitem[Doroshkevich et al.(2001)]{Doroshkevich01}
Doroshkevich, A. G., et al. 2001, \mnras, 322, 369

\bibitem[Dressler(1980)]{Dressler80}
  Dressler, A. 1980, \apj, 236, 351

\bibitem[Fadda, Slezak \& Bijaoui (1998)]{FSB98}
  Fadda, D., Slezak, E., \& Bijaoui, A. 1998, \aaps, 127, 335

\bibitem[Fasano \& Franceschini(1987)]{ff87}
  Fasano, G. \& Franceschini, A. 1987, \mnras, 225, 155

\bibitem[Fazio et al.(2004)]{Fazio04}
Fazio, G. G., et al. 2004, \apjs, 154, 10

\bibitem[Fujita(2004)]{Fujita04}
  Fujita, Y. 2004, PASJ, 56, 29

\bibitem[Geach et al.(2006)]{Geach06}
  Geach, J. E., et al. 2006, \apj, 649, 661

\bibitem[Gerken et al.(2004)]{Gerken04}
  Gerken, B.,  et al. 2004, \aap, 421, 59

\bibitem[Girardi et al.(1996)]{Girardi96}
  Girardi, M.,  et al. 1996, \apj, 457, 61

\bibitem[Goto et al.(2003)]{Goto03}
  Goto, T., Yamauchi, C., Fujita, Y., et al. 2003, \mnras, 346, 601

\bibitem[Kennicutt(1998)]{Kennicutt98}
  Kennicutt, Jr., R. C. 1998, \araa, 36, 189

\bibitem[Kodama et al.(2004)]{Kodama04}
  Kodama, T., et al. 2004, \mnras, 354, 1103

\bibitem[Lacy et al. (2007)]{Lacy07}
  Lacy, M. et al. 2007, \aj, 133, 186

\bibitem[Lima-Neto \& Durret(2007)]{LND07}
  Lima-Neto, G. \& Durret, F. 2007, \aap, submitted

\bibitem[Lin, Mohr \& Stanford(2003)]{LMS03}
  Lin, Y.-T., Mohr, J. J., \& Stanford, S. A. 2003, \apj, 591, 749

\bibitem[Mamon(1996)]{Mamon96}
  Mamon, G. 1996, in \emph{Third Paris Cosmology Colloquium}, H. J.
  de Vega \& N. S´anchez eds., p. 95, arXiv:astro-ph/9511101

\bibitem[Marcillac et al.(2007)]{Marcillac07}
  Marcillac, D., et al. M. 2007, \apj, 654, 825

\bibitem[Metcalfe, Fadda \& Biviano (2005)]{MFB05}
  Metcalfe, L., Fadda, D., \& Biviano, A. 2005, \ssr, 119, 425

\bibitem[Navarro, Frenk \& White (2005)]{NFW97}
  Navarro, J. F., Frenk, C. S., \& White, S. D. M. 1997, \apj, 490, 493

\bibitem[Pimbblet, Drinkwater \& Hawkrigg(2004)]{PDH04}
  Pimbblet, K. A., Drinkwater, M. J., \& Hawkrigg, M. C. 2004, \mnras, 354, L61

\bibitem[Polletta et al.(2007)]{Polletta07}
  Polletta, M.,  et al. 2007, \apj, 663, 81

\bibitem[Porter \& Raychaudhury(2007)]{PR07}
  Porter, S. C. \& Raychaudhury, S. 2007, \mnras, 375, 1409

\bibitem[Prokhorov \& Durret (2007)]{PD07}
  Prokhorov, D. A.\& Durret, F. 2007, \aap, in press (preprint: arXiv:0708.2166)

\bibitem[Rieke et al.(2004)]{Rieke04}
  Rieke, G. H.,  et al. 2004, \apjs, 154, 25

\bibitem[Rines et al.(2005)]{Rines05}
  Rines, K., et al. 2005, \aj, 130, 1482

\bibitem[Silva et al.(1998)]{Silva98}
  Silva, L., et al. 1998, \apj, 509, 103

\bibitem[Skrutskie et al.(2006)]{Skrutskie06}
  Skrutskie, M. F.,  et al. 2006, \aj, 131, 1163

\bibitem[Treu et al.(2003)]{Treu03}
  Treu, T.,  et al. 2003, \apj, 591, 53

\bibitem[Werner et al.(2004)]{Werner04}
  Werner, M. W.,  et al. 2004, \apjs, 154, 1

\bibitem[Young \& Scoville(1991)]{YS91}
  Young, J. S. \& Scoville, N. Z. 1991, \araa, 29, 581

\bibitem[Zabludoff \& Franx(1993)]{ZF93}
  Zabludoff, A. I. \& Franx, M. 1993, \aj, 106, 1314 

\end{thebibliography}
